\documentclass{sig-alternate-05-2015}

\usepackage{etoolbox}
\makeatletter
\patchcmd{\maketitle}{\@copyrightspace}{}{}{}
\makeatother

\usepackage{amsmath}
\usepackage{tikz}
\usetikzlibrary{positioning}
\usetikzlibrary{decorations.pathreplacing}

\usepackage{natbib}
\begin{document}



\conferenceinfo{2016 WSDM Cup Challenge}{February 22--25, 2016, San Francisco, CA, USA}


%
\conferenceinfo{Web Search and Data Mining}{'16 San Francisco, CA, USA}

\title{Simple Yet Effective Methods for Large-Scale Scholarly Publication Ranking}
\subtitle{KMi and Mendeley (team BletchleyPark) at WSDM Cup 2016}

%
%
%
%
%

\numberofauthors{2} 
%
\author{
%
%
\alignauthor
Drahomira Herrmannova\\
       \affaddr{Knowledge Media Institute, The Open University}\\
       \affaddr{Milton Keynes, United Kingdom}\\
       \email{d.herrmannova@open.ac.uk}
\alignauthor
Petr Knoth\\
       \affaddr{Mendeley Ltd.}\\
       \affaddr{London, United Kingdom}\\
       \email{petr.knoth@mendeley.com}
}
\date{\today}

\maketitle
\begin{abstract}
With the growing amount of published research, automatic evaluation of scholarly publications is becoming an important task. In this paper we address this problem and present a simple and transparent approach for evaluating the importance of scholarly publications. Our method has been ranked among the top performers in the WSDM Cup 2016 Challenge. The first part of this paper describes our method. In the second part we present potential improvements to the method and analyse the evaluation setup which was provided during the challenge. Finally, we discuss future challenges in automatic evaluation of papers including the use of full-texts based evaluation methods.
\end{abstract}

%
%


%
%

%
%



\section{Introduction}

Finding important and influential scholarly work is an essential part of any research activity as well as the cornerstone of research evaluation. However, manual discovery and evaluation of scholarly publications is with the increasing amount of published work becoming nearly impossible. This makes creating algorithms, metrics and software tools for automatic scholarly publication evaluation and ranking an important task. This is also the aim of the 2016 WSDM Cup, in which the challenge is to assess the query-independent importance of scholarly articles. The performance of different methods is evaluated on the Microsoft Academic Graph\footnote{\url{http://research.microsoft.com/en-us/projects/mag/}} (MAG) \cite{Sinha2015}, a large heterogeneous graph modelling real-life academic communication.

This paper is organised as follows. In section \ref{sec:method}, we describe the dataset and present our approach to assessing the importance of scholarly articles. In section \ref{sec:evaluation}, we report the performance of our method and propose potential improvements. Section \ref{sec:duscussion} is dedicated to the analysis of the task and discusses the applicability and benefits of new emerging approaches for solving this task. We summarise our findings in Section \ref{sec:conclusion}.


\section{Publication Ranking Methods}
\label{sec:method}
\subsection{The task and the data}
The 2016 WSDM Cup Challenge can be described as follows: given a heterogeneous graph, which models real-life academic communication, find a static rank value for each publication entity in the graph representing the papers' importance in the graph. Our approach to solving this task is in detail described in the remainder of this section.

In the 2016 WSDM Cup Challenge the performance of different methods is assessed on the MAG graph, which consists of six types of entities: scholarly publications, authors, institutions, fields of study, venues (journals and conferences, e.g. WSDM) and events (specific conference instances, e.g. WSDM 2016). The dataset also contains citation relationships between the publication entities. The relationships between the MAG graph entities are described in \cite{Sinha2015}.


\subsection{Our approach}

Our approach is based on the hypothesis that the importance of a publication can be determined by a mixture of factors evidencing its impact and the importance of entities which participated in the publication's creation. We believe method transparency is an important characteristic, for this reason we were trying to come up with a simple, understandable and transparent method which could potentially improve the current situation in research evaluation. The approach used in our submission is based on the following method. We separately score each of the types of entities in the graph (we produce a separate score for authors, institutions, journals, etc.). We then use the separate scores to provide a publication score (e.g. we score publications based on the scores of their authors, or based on the venue at which they were published). In this way we produce several different scores for the publication entities. The final score, which determines the publication's rank among its peers, is then calculated using linear combination of these scores. The standard approach for determining weights for the separate scores would be to use machine-leaning approach, however because no ground truth data were available for training and verifying the methods, we deduced the weights experimentally. Equation \ref{eq:all} shows the final weights. This equation was used to produce our final submission in the second round of the challenge.

\begin{equation}
    \begin{split}
        score(p) = &2.5 \cdot s_{pub} + 0.1 \cdot s_{age} + 1.0 \cdot s_{pr} + \\ 
                  &1.0 \cdot s_{auth} + 0.1 \cdot s_{venue} + 0.01 \cdot s_{inst} 
    \end{split}
\label{eq:all}
\end{equation}


The differences between our first and second round submissions, each of the separate ranks as well as which alternatives did we experiment with are described in the remainder of this section.

\subsection{Publication-based scoring functions}

To score the publication entities directly, without considering the score or importance of their authors or venues, we have utilised the citation relationships provided in the graph. The simplest option is to score the publications solely by the number of citations they receive. We have experimented with several options of normalising and weighting the citations, namely:

\textit{Applying a time decay to citations.} We have used an exponential decay function $f(t) = e^{-\alpha(t_{c}-t)}$, where $t_c$ is the current year, $t$ is the year in which the paper from which the citation originates was published and $\alpha$ is a constant influencing the decay rate. This means that each citation contributes to the total fully only in the year in which it originates, and the value of the citation diminishes with age. The rationale behind this is to distinguish between publications which received attention only years after publication and those which are still presently used \cite{DelCorso2009}. We have experimented with several different values of $\alpha$.

\textit{Applying a decay function to total citation counts.} The idea behind applying a decay function to the citation total is that the importance of publications doesn't necessarily increase linearly with the increasing number of received citations. For example, it has been suggested that the concept called the \textit{Matthew effect}, where highly cited papers (as well as researchers, etc.) receive a cumulative advantage, could be at work in science \cite{Merton1968,Price1976}. We have experimented in using logarithmic and linear decay, however we have achieved the best results when simply setting a maximum threshold for the total citation count above which the received citations are no longer considered.

\textit{Using mean citation counts.} Normalising total citations to citations received per year since the publication of the paper, per author of the paper, and per year and author. It has been suggested that the number of authors on the paper could cause a multiplication effect of specific audiences for each involved author \cite{Bornmann2014}. The use of citations per year is a simplification of the time decay function.

We have found the total number of citations per author of the publication with maximum threshold for the citation total to perform the best. We write this part of the equation as follows:

\begin{equation}
s_{pub}(p) = 
\begin{cases}
    c(p) / |A_p|, & \text{for } c(p) \leq t\\
    t / |A_p|, & \text{for } c(p) > t
\end{cases}
\label{eq:pub}
\end{equation}

where $c(p)$ is the total number of citations received by $p$, $A_p$ is the set of authors of $p$ and $t$ is the threshold. We have experimentally set the threshold to $t=5000$. This version of the equation is a slightly updated version for the second round of the challenge. In the first round, the second part of the equation was defined as $0 / |A_p|, \text{for } c(p) > t$.

Furthermore, to account for publication age, we use a score based on the age. This score is a simple linear function of publication year and can be written as 

\begin{equation}
s_{age}(p) = y_p
\end{equation}

where $y_p$ is the year of publication of $p$. Based on this score, papers published in the current year have the highest importance and as time elapses their importance linearly decreases.


In the second phase of the WSDM Cup Challenge we have also computed the PageRank \cite{Brin1998} value for each of the publication entities in the graph. To allow for efficient PageRank calculation, we chose an approach similar to \cite{Bini2008} and introduced a new ``dummy'' paper in the network, which is cited and cites all publications in the citation network except for itself. This paper collects and redistributes weight equally to all publications in the network. This part of the equation can be written as

\begin{equation}
s_{pr}(p) = PR(p)
\end{equation}

We have found the PageRank score to perform similarly to total citation counts and we added the PageRank value as an additional feature.

\subsection{Author-based score}

Commonly used methods for evaluating author performance include the total number of citations received by an author, average number of citations per author's publication and indices such as the h-index \cite{Hirsch2005}. We have experimented with these three methods. We calculated the given value for each of the authors of a publication and then tested ranking the publication entities using the maximum, total and mean of the values of the publication's authors (e.g. using maximum, total and mean of the authors' h-index values). We found the mean value of citations per author's publication to perform the best. The author-based rank we used can then be expressed as

\begin{equation}
s_{auth}(p) = \frac{\sum_{a \in A_p}\frac{\sum_{x \in P_a}c(x)}{|P_{a}|}}{|A_{p}|}
\end{equation}

where $P_a$ is a set of publications authored by $a$.


\subsection{Venue-based score}

The metric which is considered the standard in journal evaluation is the Journal Impact Factor (JIF) \cite{Garfield1955}. The JIF calculation concerns the computation of a mean number of citations received per item published in the journal during a specified time frame, typically during two years prior to the current year. Alternative journal evaluation metrics include the Scimago Journal Rank\footnote{http://www.scimagojr.com} and the Eigenfactor \cite{Bergstrom2007} which both revolve around the idea that citations from high-impact journals provide a larger contribution to the importance of a journal than citations from poorly ranked journals.

In evaluating conferences no established metric similar to JIF or other journal evaluation metrics exists. However, a similar approach as in case of journals can be used also for evaluating conferences. We have experimented with few simple scoring functions, such as with total number of citations received by a venue and mean number of citations per paper published at the venue, and with applying these scores to the papers published at the venue (this is an approach similar to the JIF, however we have used all papers published during the existence of the journal or conference). Our final venue-based score can be calculated as

\begin{equation}
s_{venue}(p) = \sum_{x \in P_v, x \neq p}c(x)
\end{equation}

where $P_v$ is a set of papers published at a venue $v$.


\subsection{Institution-based score}

Various approaches exist to evaluating institutions. The Nature publishing group ranks institutions based on the number of articles published in their journal Nature\footnote{http://www.natureasia.com/en/publishing-index/global/}. Scimago Institution Rankings\footnote{www.scimagoir.com/} provide a list of indicators, including the total number of documents published in scholarly journals, proportion of highly cited publications and rate of collaboration with foreign institutions. In our approach we have however used a simple method similar to the author and venue score. Our final institution-based score can be expressed as 

\begin{equation}
s_{inst}(p) = \frac{\sum_{i \in I_p} \sum_{x \in P_i, x \neq p} c(x)}{|I_p|}
\end{equation}

where $I_p$ is a set of (unique) institutions of the authors of the publication and $P_i$ is a set of publications published by authors affiliated with institution $i$.


\section{Experiments}
\label{sec:evaluation}

\subsection{Evaluation}

Details of the evaluation dataset and metric were not provided. According to the organisers the submitted results were evaluated based on the percentage agreements with human evaluation data \cite{WsdmRules}. The evaluation data were prepared by Computer Science experts who conducted pairwise ranking of a subset of the MAG dataset. The evaluation data have then been split into validation and test set. While the challenge was running, the participants could evaluate their results against the test data through an online evaluation tool, which provided a score for each of the submitted runs. At the end of the first round of the challenge, the last submitted run of each team was scored against the validation set. During the training phase of the challenge we have submitted over 270 runs.

\subsection{Performance comparison with other teams}

The performance of all participating teams was provided both during and after the first round of the challenge through a public leaderboard. According to the leaderboard ranks, our method has achieved the highest score on the test data and has been ranked as fifth best when scored against the validation data.

\subsection{Potential improvements}
\label{sec:improvements}

There is a number of ways in which our method could be improved. We believe the main possibilities include the following options.

\textit{Better utilisation of the citation network.} Due to resource limitations, we were only able to compute PageRank of the publication entities later in the challenge. We see a potential improvement in computing additional network measures, such as different centrality indices, for all entities in the graph.

\textit{Inclusion of additional data sources.} At the beginning of the challenge we explored the possibility of obtaining additional data. In particular we were interested in utilising altmetric \cite{Galligan2013} and webometric \cite{Almind1997} data sources and acquiring publication full-texts or abstracts for use in semantometric measures \cite{Knoth2014}. For altmetric and webometric data we have investigated the feasibility of obtaining data from Altmetric.com, Mendeley, ResearchGate, ImpactStory and ArXiv. For the publication full-texts we have investigated Elsevier, Springer, CrossRef and Mendeley APIs. Unfortunately most of the investigated services either didn't provide an interface for downloading all of their data, or their coverage was too low, which is why we eventually dropped this idea. However, particularly if access to the publication full-texts was possible, this option could provide valuable additional information, for example by extending simple citation counts to research contribution \cite{Knoth2014}. A more detailed discussion of the alternative methods is provided in Section \ref{sec:alt}.

\textit{Possibility to analyse the evaluation data and metric.} At the moment it is not clear if and up to what extent do the expert judgements correspond with the importance of the publications. Publishing the evaluation dataset and the metric would help in understanding whether the methods submitted to the challenge could help in improving user experience and research evaluation.

\textit{Revise the maximum citation threshold} used the $s_{pub}$ score. It is yet to be determined why this threshold led to the improvement of our results.


\section{Discussion}
\label{sec:duscussion}

\subsection{What have we learned}
\label{sec:learned}
In scoring each of the graph entities we have experimented with different options, from simple citation counts to applying decay functions, calculating PageRank and h-index. It is interesting that in each case, a method based on simple citation counts produced better results than using these widely used measures. Regardless of whether better scoring functions can be found, we believe that in order to develop a more optimal ranking method, it is crucial to better understand the evaluation data and method (what is required from the ranking system). Although a simple approach based on citation counts produced the best results, this doesn't mean such method will work equally well in real-life settings. For example, it is not clear how much are the human judgement data biased towards citation counts. This issue could manifest in case the judges had access to such information when rating the publications. Furthermore, although citation counting provides a simple and easily understandable ranking method, it does not account for many characteristics of citations, including the differences in their meaning \cite{Nicolaisen2007}, popularity of certain topics and types of research papers \cite{Seglen1997}, the skewness of the citation distribution \cite{Seglen1992} and the time delay for citations to show up \cite{Priem2010}.

\subsection{Evaluation}

The goal of the 2016 WSDM Cup challenge was to assess the importance of scholarly articles while exploring alternatives to citations, which suffer from many drawbacks (Section \ref{sec:learned}). The format of the results is in WSDM'16 Cup defined as ranked list of the MAG publication entities. In order to evaluate these results, the evaluation setup consisted of the evaluation data -- reference ranks prepared by human judges -- and an evaluation metric. While preparing our submission, we have identified few problems of the evaluation setup. One of these problems, which we discussed in Section \ref{sec:learned}, is the subjectivity of the evaluation dataset. While the description of the task encouraged exploration of approaches alternative to citations, it wasn't clear whether the evaluation setup was capable of  potentially rewarding properties of such approaches. Our citation-based method has achieved a high score. Furthermore, due to the fact that the details of the evaluation data were not shared, it became more complicated to avoid overfitting our model. The availability of a good evaluation framework is crucial for enabling the development of new ranking methods and comparison of different approaches. We believe a good evaluation framework should favor properties of the desired ranking system, and the method of creation of this dataset should be transparent to facilitate understanding any biases present in the dataset and to help preventing overfitting.

\subsection{Alternative ranking methods}
\label{sec:alt}
In section \ref{sec:improvements}, we list the external datasources which we investigated. Our motivation for exploring these external datasources was the hope of utilising new altmetric and webometric research evaluation methods. The advantage of these approaches lies for example in the early availability of the required data, when compared to the delay with which citations show up. These metrics also provide a broader view of publications' impact. However, our main interest lies in the utilisation of publication full-text for research evaluation, this set of metrics is referred to as Semantometrics \cite{Knoth2014}. In contrast to the other existing classes of metrics (Bibliometrics, Altmetrics, Webometrics, etc.) which utilise external data (typically the number of interactions in the scholarly network), Semantometrics build on the premise that the manuscript of the publication is needed to asses its value. A pilot study, which investigated the first semantometric research evaluation measure, has demonstrated the feasibility and utility of such approach \cite{Knoth2014}. The biggest problem of this approach is the difficulty of obtaining the publication full-texts, due to various copyright restrictions and paywalls. The MAG dataset could be a very valuable resource for semantometric research if it could be combined with publication full-texts, which is something we are currently investigating. An interesting future direction could be to enrich the MAG with the altmetric, webometric and semantometric data and organise another run of the challenge with the possibility to use these data.

\section{Conclusions}
\label{sec:conclusion}

In this paper we presented our method for assessing the importance of scholarly publications, which we submitted to the 2016 WSDM Cup Challenge. Our method was ranked among the top performers in the challenge. We have presented several potential improvements to the method and the knowledge acquired when carrying out experiments. Our findings highlight the difficulty of progressing beyond citation counts. While MAG is an extremely useful dataset for testing evaluation metrics, we need this dataset to be merged with other sources evidencing impact, including data required by Webometrics and Semantometrics, to develop and test fundamentally new metrics. Additionally, there is a need for a large, open and unbiased dataset of human judgements to move us closer to this goal. 


%
\bibliographystyle{abbrv}
\bibliography{bibliography}  

\begin{thebibliography}{10}

\bibitem{WsdmRules}
{WSDM Cup Ranker Challenge Rules}.
\newblock \url{https://wsdmcupchallenge.azurewebsites.net/Home/Rules}.
\newblock Accessed: 2015-12-14.

\bibitem{Almind1997}
T.~C. Almind and P.~Ingwersen.
\newblock {Informetric analyses on the world wide web: methodological
  approaches to 'webometrics'}.
\newblock {\em Journal of Documentation}, 53(4):404--426, 1997.

\bibitem{Bergstrom2007}
C.~Bergstrom.
\newblock {Eigenfactor: Measuring the value and prestige of scholarly
  journals}.
\newblock {\em C{\&}RL News}, 68(5):314--316, 2007.

\bibitem{Bini2008}
D.~a. Bini, G.~M. {Del Corso}, and F.~Romani.
\newblock {Evaluating Scientific Products by Means of Citation-Based Models: A
  First Analysis and Validation}.
\newblock {\em Electronic Transactions on Numerical Analysis}, 33:1--16, 2008.

\bibitem{Bornmann2014}
L.~Bornmann and L.~Leydesdorff.
\newblock {Does Quality Matter for Citedness? A comparison with para-textual
  factors and over time.}, 2014.

\bibitem{Brin1998}
S.~Brin and L.~Page.
\newblock {The Anatomy of a Large-Scale Hypertextual Web Search Engine}.
\newblock {\em Proceedings of the 7th International World-Wide Web Conference},
  page~20, 1998.

\bibitem{DelCorso2009}
G.~M. {Del Corso} and F.~Romani.
\newblock {A time-aware citation-based model for evaluating scientific
  products: extended abstract}.
\newblock {\em Proceedings of the 4th International ICST Conference on
  Performance Evaluation Methodologies and Tools}, (October), 2009.

\bibitem{Galligan2013}
F.~Galligan and S.~Dyas-Correia.
\newblock {Altmetrics: Rethinking the Way We Measure}.
\newblock {\em Serials Review}, 39(1):56--61, mar 2013.

\bibitem{Garfield1955}
E.~Garfield.
\newblock {Citation indexes for science. A new dimension in documentation
  through association of ideas}.
\newblock {\em Science}, 122(3159):108--11, oct 1955.

\bibitem{Hirsch2005}
J.~E. Hirsch.
\newblock {An index to quantify an individual's scientific research output}.
\newblock {\em Proceedings of the National Academy of Sciences of the United
  States of America}, 102(46):16569--72, nov 2005.

\bibitem{Knoth2014}
P.~Knoth and D.~Herrmannova.
\newblock {Towards Semantometrics: A New Semantic Similarity Based Measure for
  Assessing a Research Publication's Contribution}.
\newblock {\em D-Lib Magazine}, 20(11/12), 2014.

\bibitem{Merton1968}
R.~K. Merton.
\newblock {The Matthew Effect in Science}.
\newblock {\em Science}, 159(3810):56--63, 1968.

\bibitem{Nicolaisen2007}
J.~Nicolaisen.
\newblock {Citation Analysis}.
\newblock {\em Annual Review of Information Science and Technology}, 41(1),
  2007.

\bibitem{Price1976}
D.~J. d.~S. Price.
\newblock {A General Theory of Bibliometric and Other Cumulative Advantage
  Processes}.
\newblock {\em Journal of the American Society for Information Science},
  27(5-6):292--306, 1976.

\bibitem{Priem2010}
J.~Priem and B.~M. Hemminger.
\newblock {Scientometrics 2.0: Toward new metrics of scholarly impact on the
  social Web}.
\newblock {\em First Monday}, 15(7), jul 2010.

\bibitem{Seglen1992}
P.~O. Seglen.
\newblock {The Skewness of Science}.
\newblock {\em Journal of the American Society for Information Science},
  43(9):628--638, oct 1992.

\bibitem{Seglen1997}
P.~O. Seglen.
\newblock {Why the impact factor of journals should not be used for evaluating
  research}.
\newblock {\em BMJ: British Medical Journal}, 314(February):498--502, 1997.

\bibitem{Sinha2015}
A.~Sinha, Z.~Shen, Y.~Song, H.~Ma, D.~Eide, B.-j.~P. Hsu, and K.~Wang.
\newblock {An Overview of Microsoft Academic Service (MAS) and Applications}.
\newblock In {\em Proceedings of the 24th International Conference on World
  Wide Web}, pages 243--246, Florence, Italy, 2015. ACM Press.

\end{thebibliography}
%
%
\end{document}